# Phase Coexistence in $Cs_3Bi_2I_9$ Ferroelastics: Optical, Dilatation and Ultrasonic Velocity Studies

[1]I.Girnyk, [1]O.Krupych, [1]I.Martunyuk-Lototska, [2]F.Motsnyi and [1]R.Vlokh

[1]Institute of Physical Optics, 23 Dragomanov Str., 79005, Lviv, Ukraine, e-mail: vlokh@ifo.lviv.ua
[2]V.Lashkaryav Institute of Semiconductor Physics of NAS of Ukraine, 41 Nauky Av., 03028, Kiev, Ukraine, e-mail: motsnui@sun.semicond.kiev.ua

## Abstract

The results for dilatometric, ultrasonic velocity and the domain structure studies of $Cs_3Bi_2I_9$ crystals in the course of ferroelastic phase transition are presented. It is shown that the phase transition in $Cs_3Bi_2I_9$ is of the first order. Basing on the thermal expansion results and the data obtained for the elastic stiffness of $Cs_3Bi_2I_9$ crystals, the latent heat $Q$=79J/kg at $T_c$ is calculated. The region between 183K and 221K (at cooling) is identified as a heterostructure region, where ferroelastic and paraelastic phases coexist.

**Key words**: ferroelastics, domain structure, heterophase structure, ultrasonic velocity, latent heat, $Cs_3Bi_2I_9$ crystals

## Introduction

$Cs_3Bi_2I_9$ crystals belong to a large family of layered-structure compounds with a general formula $A_3B_2X_9$, where A=Cs, Rb, Tl, $NH_4$ and K; B=Fe, As, Sb, Bi, Cr, W and Mo, and X=I, Br and Cl. These crystals possess the point group of symmetry 6/mmm at the room temperature [1]. $Cs_3Bi_2I_9$ undergoes the structural phase transition with the symmetry change of 6/mmm↔2/m at $T_c$=220K [2]. According to the known *K.Aizu's* classification [3], this phase transition should be ferroelastic. On the basis of NQR study it has been concluded in [4] that the intermediate incommensurate phase exists in $Cs_3Bi_2I_9$ crystals below $T_c$=220K. On the other side, the X-ray [2] and neutron diffraction [5] studies have not confirmed the existence of incommensurately modulated phase and the authors of [2] have assumed a possibility of phase coexistence in the temperature range 205K-220K. Besides, the phase coexistence concept contradicts the order of phase transition determined for $Cs_3Bi_2I_9$. Using the studies of the macroscopic properties (the birefringence, elastic stiffness and the thermal expansion), the authors [6] have concluded that the phase transition in $Cs_3Bi_2I_9$ is of the second order. This is why the aim of the present study is to clarify the nature of the intermediate phase in $Cs_3Bi_2I_9$ crystals.

## Experimental

The investigation of the phase transition in $Cs_3Bi_2I_9$ crystals was carried out with the ultrasonic velocity and dilatometric methods, as well as the observation of domain structure by means of optical polarization microscopy. The temperature dependences of the ultrasonic wave velocities have been measured with the pulse-echo overlap method [7]. The accuracy of the absolute velocity determination was about 0.5%. The acoustic waves in samples were excited with $LiNbO_3$ transducers, characterized with the resonance frequency $f$ = 10 MHz, the bandwidth $\Delta f$ = 0.1 MHz and the acoustic power $P_a$ = 1 to 2W. The thermal expansion was measured with the capacity dilatometer, while the domain structure observations were performed using the polarization microscope.

## Results and discussion

The temperature dependences of the velocities of transverse and longitudinal acoustic waves propagated along the $c$-axis and polarized parallel to <101> and <001> directions are presented in Figure 1. The elastic stiffness coefficients have been calculated from the acoustic wave velocities and the crystal density value $\rho$=5.02×10$^3$kg/m$^3$ taken from [6] (see Figure 1). As seen from Figure 1, both the velocity of the transverse acoustic wave and the elastic stiffness coefficient $C_{55}$ approach zero at $T_c$=221K. In the temperature range below 221K, the measurements of the acoustic wave velocity become complicated due to a prominent ultrasonic wave scattering by the phase and domain boundaries. In the heating run, the reflected acoustic signal has revived again at $T$=226K. The acoustic velocity at this temperature is equal to $v_{13}$=240m/s. Such the temperature behaviour of the elastic properties in Cs$_3$Bi$_2$I$_9$ crystals can be explained by the proper character of ferroelastic phase transition, which is accompanied with appearing the spontaneous strain component $e_{31}$ at $T_c$. Contrary to the $C_{55}$ coefficient, the $C_{33}$ elastic stiffness does not exhibit softening at $T_c$.

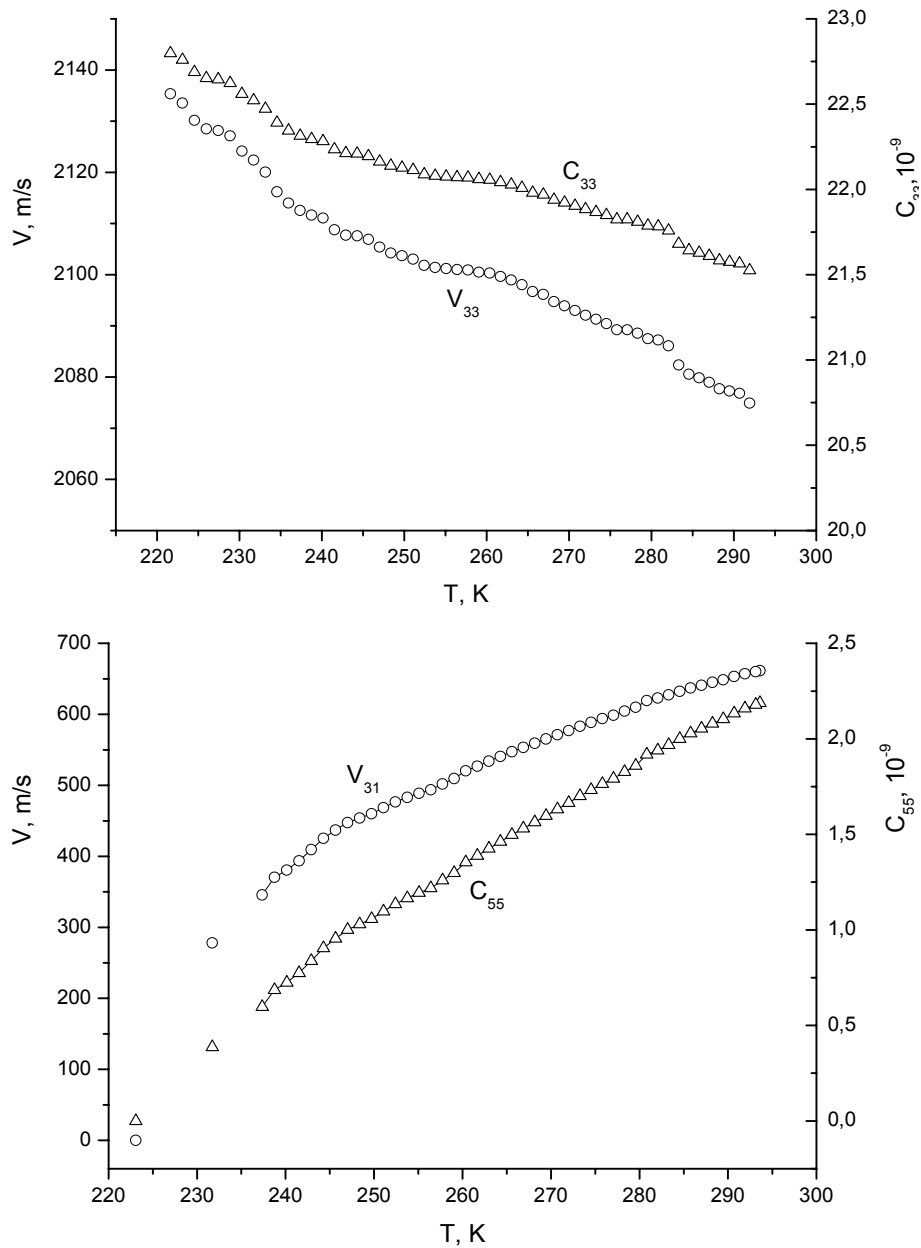

**Fig. 1.** Temperature dependences of the elastic stiffness and the ultrasonic velocities for Cs$_3$Bi$_2$I$_9$ crystals.

The temperature dependences of the thermal expansion measured along the $c$-axis are presented in Figure 2.

One can see that the behaviour of thermal expansion at $T_c$ is different in the heating and cooling runs. At heating, the peak at $T_c$ is observed. This peak is associated with the expansion of sample due to absorption of the first-order phase transition latent heat at $T_c$. The absence of similar anomalous dependence in the thermal expansion at cooling is caused by masking this decrement by the spontaneous deformation, which leads to compressing of sample. Moreover, the $T_c$ temperature values at heating and cooling do not coincide with each other. The first-order phase transition temperature hysteresis is equal to $\Delta T_h$=5K. The equation for the latent heat may be written as

$$Q_L = T_c \alpha_\mu \Delta\sigma_\mu = T_c \alpha_\mu C_{\mu\mu} \Delta e_\mu, \quad (1)$$

where $\alpha_\mu$ is the tensor of thermal expansion coefficients, $\Delta\sigma_\mu$ and $\Delta e_\mu$ the increments of mechanical stress and strain appearing owing to the absorption or emission of the latent heat. The determined values of the strain increment, the thermal expansion coefficient and the elastic stiffness are respectively $\Delta e_\mu$=0.46×10$^{-3}$, $\alpha_3$=1.7×10$^{-4}$ and $C_{33}$=2.28×10$^{10}$N/m$^2$. Thus, the calculated value of the latent heat at $T_c$ turns out to be $Q_L$=79J/kg. A relatively small value of the latent heat means that the first-order ferroelastic phase transition in Cs$_3$Bi$_2$I$_9$ crystals is close to the second-order. It is interesting to note that the temperature dependence of the strain is quite unusual: the spontaneous strain (i.e., the difference between the values of thermal expansion extrapolated from the paraelastic phase and the actual thermal expansion in the ferroelastic phase) changes its sign approximately at $T\approx$125K. The latter could be related to the existence of additional phase transition somewhere near this temperature point.

At $T_c$=221K in the cooling run, we have observed the appearance of heterophase structure in Cs$_3$Bi$_2$I$_9$ crystals, which consists of the regions of multidomain ferroelastic phase and the regions of paraelastic phase. The phases are separated by clearly visible phase boundaries (see Figure 3,a). The shape of the phase boundaries is the curved surfaces perpendicular to the (001)-plane. Following [8], we remind that the elastic strains appearing at the first-order phase transition with the symmetry change of $6/mmm \leftrightarrow 2/m$ are not in general compatible on the phase boundary. According to the approximate (i.e., derived with assuming $e_{13} \approx 0$) solution of the elastic compatibility equation under the condition of matching the single-domain ferroelastic phase with the paraelastic phase, the traces of the phase boundaries on (001)-plane should be characterized with $x = \pm y$ orientation. The distortion of the phase boundary from a planar shape is most probably connected with the fact that $e_{13} \neq 0$ and the phase matching takes place, owing to a compensation of macroscopic strains on the phase boundary by a spontaneous compilation of domain configuration [9]. The heterophase structure gradually disappears with temperature decreasing and exists down to 183K. The crystal becomes multidomain, though single-phase, below this temperature (see Figure 3,b). At heating, the heterophase structure also exists in nearly the same temperature region, but it differs from that occurring at cooling. In particular, we have observed at least three pairs of mutually perpendicular domain walls W and W′ (the notation being according to *J.Sapriel's* classification [10]) in the ferroelastic phase (see Figure 3, b and c).

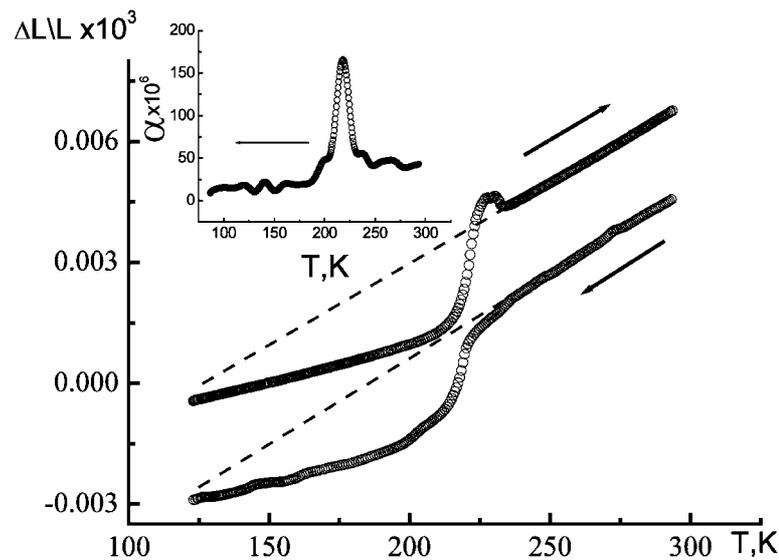

**Fig. 2.** Temperature dependences of the thermal expansion for Cs$_3$Bi$_2$I$_9$ crystals along the $c$-axis at heating and cooling. The insert shows the temperature dependence of thermal expansion coefficient.

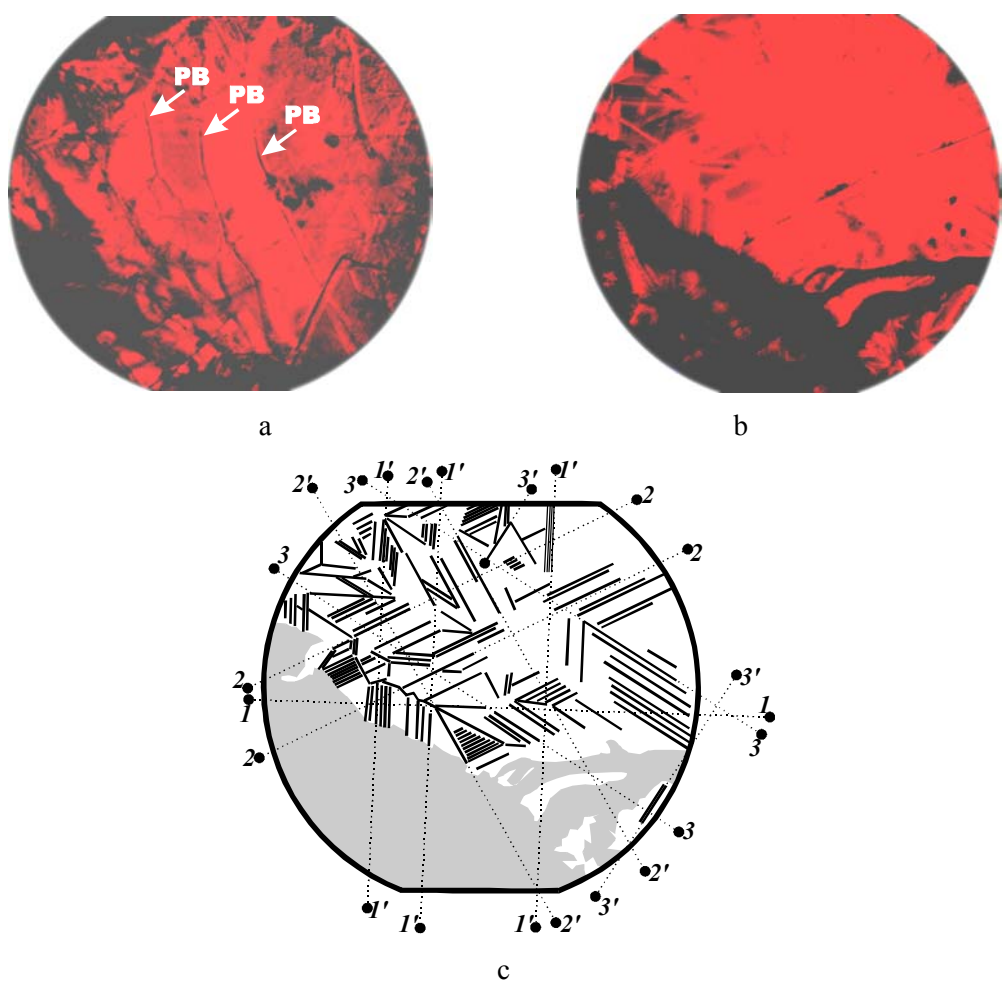

**Fig. 3.** (a) The heterophase at T=213K, (b) the domain structure at *T*=181K observed in $Cs_3Bi_2I_9$ crystals along the *c*-axis, and (c) the schematic view of the domain structure. PB denotes the phase boundary, while 11 and 1′1′, 22 and 2′2′, 33 and 3′3′ the pairs of mutually perpendicular W and W′ domain walls.

## Conclusions

In the present paper the dilatometric and ultrasonic velocity studies, as well as the microscopic observations, have been performed for Cs$_3$Bi$_2$I$_9$ crystals in the region of their ferroelastic phase transition. It has been shown that the phase transition is of the first order. On the basis of thermal expansion measurements for Cs$_3$Bi$_2$I$_9$ and the corresponding data obtained for the elastic stiffness, the value of the latent heat at $T_c$ has been calculated as $Q$=79J/kg. The region from 183K to 221K (at cooling) has been identified as the heterophase region, where the ferroelastic and paraelastic phases coexist. The ferroelastic domain structure existing in Cs$_3$Bi$_2$I$_9$ crystals below 183K corresponds to the change of point symmetry 6/mmm$\leftrightarrow$2/m .

## Acknowledgement

The authors [1] are grateful to the Ministry of Education and Science of Ukraine for the financial support of the present study under the Projects N0101U007216 and 0103U000703.